\documentclass[epjCONF]{svjour}
\usepackage{graphics}
\usepackage[varg]{txfonts} 
\usepackage[latin1]{inputenc}
\session-title{Tidal Disruption events and AGN outbursts workshop}
\begin{document}
\title{Tidal disruption flares from stars on eccentric orbits}
\author{Kimitake Hayasaki\inst{1,2,3}\fnmsep\thanks{\email{kimi@kusastro.kyoto-u.ac.jp; kimi@kasi.re.kr}} 
\and Nicholas Stone\inst{2} \and Abraham Loeb\inst{2}}
\institute{Department of Astronomy, Kyoto University, Kitashirakawa-Oiwake-cho, Sakyo-ku, Kyoto 606-8502, Japan
\and Harvard-Smithsonian Center for Astrophysics, 60 Garden Street, Cambridge, MA, 02138, USA
\and Korea Astronomy and Space Science Institute, Daedeokdaero 776, Yuseong, Daejeon 305-348, Korea}
\abstract{
We study tidal disruption and subsequent mass fallback
for stars approaching supermassive black holes 
on bound orbits, by performing
three dimensional Smoothed Particle Hydrodynamics simulations 
with a pseudo-Newtonian potential. 
We find that the mass fallback rate decays with the expected -5/3 power of time
for parabolic orbits, albeit with a slight deviation due to the self-gravity of the stellar debris.
For eccentric orbits, however, there is a critical value of the orbital eccentricity, significantly below which all of the stellar debris is bound to the supermassive black hole.  All the mass therefore falls back to the supermassive black hole in a much shorter time than in the standard, parabolic case. The resultant mass fallback rate considerably exceeds
the Eddington accretion rate and substantially differs from the -5/3 power of time.
} 
\maketitle
%
\section{Introduction}
\label{sec:intro}
There is substantial evidence that galactic nuclei harbor 
supermassive black holes (SMBHs), the majority of which are 
quiescent and not active galactic nuclei.
The tidal disruption of a star by a SMBH, and subsequent flaring activity, 
provides a rare observational diagnosis for the large population of quiescent SMBHs. 
These powerful flares are expected to have a luminosity at least comparable 
to the Eddington luminosity \cite{rmj88,sq09}.

The standard picture of a tidal disruption event (TDE) involves a star at
large separation falling into a massive black hole on an almost parabolic orbit. 
After the star is tidally disrupted by the SMBH, half the stellar debris becomes gravitationally bound to the SMBH as it loses orbital energy inside the tidal radius. The bound
debris finally falls back and accretes onto the black hole. Kepler's third
law implies that the accretion rate decays with the -5/3 power of time \cite{rmj88,ecr89}. 

Observed light curves are in reasonable agreement with this theoretically 
predicted mass fallback rate, although some show 
deviations \cite{gs12}
and the sample size is sufficiently small to 
make detailed testing of theoretical models difficult. 
Observations suggest that the TDE rate is $\sim 10^{-5} \rm{yr}^{-1}$ per galaxy \cite{djl02}.
This observed rate is in rough agreement with theoretical rate 
estimates based on two-body scattering at $\sim \rm{pc}$ scales, which motivates the assumption of nearly parabolic orbits \cite{mt99}.

However, recent theoretical studies on rates of tidal separation of binary
stars by SMBHs suggest that a significant fraction of tidal disruption flares 
may occur from stars approaching the black hole on somewhat eccentric orbits, 
significantly less parabolic than in the standard picture \cite{asp11}. 
Other sources of TDEs from stars on more eccentric orbits include binary SMBH systems and recoiling SMBHs \cite{kh12}.   These latter two sources are capable of producing TDEs with even lower values of orbital eccentricity than in the binary separation scenario, and motivate our work here.
In this paper, we explore through hydrodynamical simulations how mass fallback rates in TDEs vary between the canonical, parabolic case and the underexplored eccentric scenario. 

%
\section{Numerical method}
\label{sec:2}
%
We describe here procedures for numerically modeling
the tidal disruption of stars on bound orbits. The simulations
presented below were performed with a three-dimensional (3D) Smoothed Particle Hydrodynamics (SPH) code,
which is based on a version originally developed by Ref. \cite{bate95}.
We model the initial star as a polytropic gas sphere in hydrostatic equilibrium. The tidal disruption
process is then simulated by setting the star in motion through the
gravitational field of an SMBH.

A star is tidally disrupted 
when the tidal force of the black hole acting on the star 
is stronger than the star's self-gravity.
The radius where these two forces balance is 
defined as the tidal disruption radius 
\begin{equation}
r_{\rm t}=\left( \frac{M_{\rm BH}}{m_*} \right)^{1/3}r_*,
\label{eq:tdr}
\end{equation}
where $M_{\rm{BH}}$ is the black hole mass and $m_*$ is the stellar mass. 
The star-black hole system is put on the $x$-$y$ plane, where both axes are normalized by $r_{\rm{t}}$ and the
black hole is put at the origin of the system. The initial position of
the star is given by $\mbox{\boldmath $r$}_0=(r_{\rm 0}\cos\phi_0,
r_{\rm 0}\sin\phi_0, 0)$, where $r_0=3r_{\rm{t}}$ 
is the radial distance from the black hole and $\phi_0$ shows the angle
between $x$-axis and $\mbox{\boldmath $r$}_0$. 
In our simulations, the black hole is represented by a sink particle
with the appropriate gravitational mass $M_{\rm{BH}}$.  All gas
particles that fall within a specified accretion radius are accreted
by the sink particle. We set the accretion radius of the black hole as
equal to the Schwarzshild radius $r_{\rm{S}}=2GM_{\rm{BH}}/c^2$, with
$c$ being the speed of light.

%
%

In order to treat approximately the relativistic precession of a test particle 
in the Schwarzschild metric, we incorporate into our SPH code 
the following pseudo-Newtonian potential \cite{wc12}:
\begin{eqnarray}
U(r)
=-\frac{GM_{\rm{BH}}}{r}\left[c_1+\frac{1-c_1}{1-c_2(r_{\rm{S}}/2r)}
+c_3\frac{r_{\rm{S}}}{2r}\right],
\label{eq:wegg}
\end{eqnarray}
where we adopt $c_1=(-4/3)(2+\sqrt{6})$, $c_2=(4\sqrt{6}-9)$, 
and $c_3=(-4/3)(2\sqrt{6-3})$. Equation~(\ref{eq:wegg}) reduces 
to the Newtonian potential when $c_1=1$ and $c_2=c_3=0$ are adopted.
Note that equation~(\ref{eq:wegg}) includes no higher-order relativistic effects
such as the black hole spin or gravitational wave emission.

We have performed five simulations of tidal disruption events 
with different parameters.
The common parameters through all of simulations are following: 
$m_*=1M_\odot$, $r_*=1R_\odot$, $M_{\rm{BH}}=10^6M_\odot$, 
$\phi_0=-0.4\pi$, and $\gamma=5/3$. 
The total number of SPH particles used in each simulation 
is $10^5$, and the termination time of each simulation is $4\Omega_{*}^{-1}$,
where $\Omega_{*}^{-1}\equiv\sqrt{r_*^3/Gm_*}
\simeq5.1\times10^{-5}(r_*/R_\odot)^{3/2}(M_\odot/m_*)^{1/2}
\,\rm{yr}$.
We also adopt standard SPH artificial viscosity parameters
$\alpha_\mathrm{SPH}=1$ or $\beta_\mathrm{SPH}=2$.
Table 1 summarizes each model, where the penetration factor $\beta$ represents 
the ratio of the tidal disruption radius to pericenter distance, $r_{\rm{p}}$.

%
\section{Tidal disruption of stars on bound orbits}
\label{sec:3}
%
As an approaching star enters into the tidal disruption radius, 
its fluid elements become dominated by the tidal force of the black hole, 
while their own self-gravity and pressure forces become relatively negligible. 
The tidal force then produces a spread in specific energy of the stellar debris
 \begin{equation}
 \Delta\epsilon\approx \frac{GM_{\rm{BH}}r_*}{r_{\rm t}^2}.
 \label{eq:spreade}
 \end{equation}
The total mass of the stellar debris is defined with the differential mass distribution 
$m(\epsilon)\equiv dM(\epsilon)/d\epsilon$, where 
$M(\epsilon)\equiv\int_{-\infty}^{\infty}m(\epsilon^{'})d\epsilon^{'}$.
When a star is disrupted from a parabolic orbit, 
$m(\epsilon)$ will be centered on zero and distributed 
over $-\Delta \epsilon\le\epsilon\le\Delta\epsilon$.

Since the stellar debris with negative specific energy 
is bound to the SMBH, it returns to pericenter and 
will eventually accrete onto the black hole. 
If we define its binding energy, $\epsilon=-GM_{\rm{BH}}/2a$ 
(the semi-major axis of the stellar debris is $a$),
then the mass fallback rate is given by \cite{ecr89}
$dM/dt=(dM(\epsilon)/d\epsilon)|d\epsilon/dt| \,\,(\epsilon<0)$, 
where $d\epsilon/dt=-(1/3)(2\pi{GM_{\rm{BH}}})^{2/3}t^{-5/3}$.
This is derived from Kepler's third law. 

%
%
\begin{table*}
 \centering
  \caption{
The first column shows each simulated scenario. 
The second, third, and fourth columns are 
the penetration factor $\beta=r_{\rm{p}}/r_{\rm{T}}$, the initial orbital 
eccentricity of star-black hole system $e_*$, and its initial semi-major axis 
$a_*$, respectively. The last column describes the remark for each model.
}
  \begin{tabular}{@{}ccccccccccl@{}}
  \hline
Model 
& {$\beta=r_{\rm{t}}/r_{\rm{p}}$} 
& {$e_*$} & {$a_*\,[r_{\rm{t}}]$} 
& {Remarks} \\
 \hline
1 &  $1$ & $1.0$  & $-$ & Newtonian \\
2 &  $1$ & $1.0$  & $-$ & Pseudo-Newtonian \\
3 &  $5$ & $1.0$  & $-$ & Pseudo-Newtonian \\
4 &  $1$ & $0.98$  & $50.0$ & Pseudo-Newtonian \\
5 &  $5$ & $0.98$  & $10.0$ & Pseudo-Newtonian \\
\hline
\end{tabular}
\end{table*}
%
%
\begin{figure}
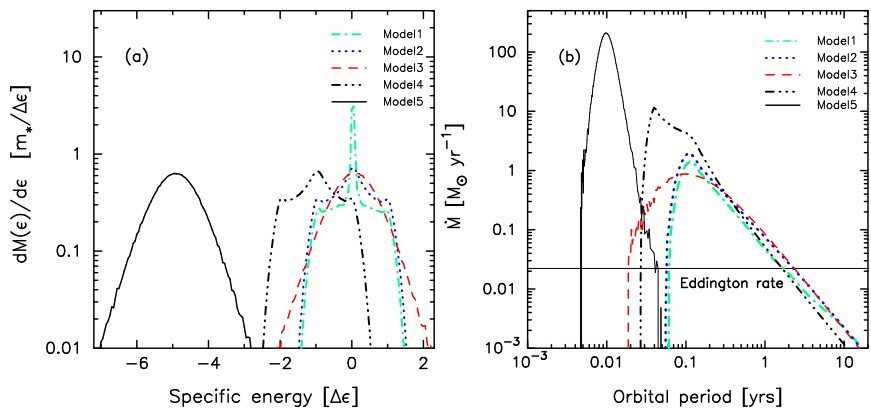

\begin{center}
\resizebox{0.8\columnwidth}{!}{
\includegraphics{fig1.ps}
\includegraphics{fig2.ps}
}
\caption{
Differential mass distributions over specific energy of stellar debris and 
their corresponding mass-fall back rates.
While the differential 
mass distribution is shown in panel (a), the mass fallback rate
is shown in panel (b). 
In both panels, the dot-dashed line (green), dotted line (blue), dashed line (red), dashed-three-dotted line (black), and solid line (black) represent the mass distributions and corresponding mass fallback rates of Model~1, Model~2, Model~3, Model~4 and Model~5, respectively. 
The energy is measured in units of $\Delta\epsilon$ given by equation~(\ref{eq:spreade}).
}
\label{fig:mass-mdot}
\end{center}
\end{figure}
%

The specific orbital energy of a star on an eccentric orbit is given by
\begin{equation}
\epsilon_{\rm{orb}}\approx-\frac{GM_{\rm{BH}}}{2a_*}
=-\frac{GM_{\rm{BH}}}{2r_{\rm{t}}}\beta(1-e_*),
\label{eq:eporb}
\end{equation}
where $a_*$ and $e_*$ are the initial semi-major axis of the star-black hole system 
and its initial orbital eccentricity, respectively.
This quantity is less than zero because of the finite
value of $a_*$, in contrast to the standard, parabolic orbit of a
star.  If $\epsilon_{\rm{orb}}$ is less than $\Delta\epsilon$, all the
stellar debris should be bounded by the black hole, even after the
tidal disruption. The condition $\epsilon_{\rm{orb}}=\Delta\epsilon$
therefore gives a critical value of orbital eccentricity of the star
\begin{equation}
e_{\rm crit}\approx1-\frac{2}{\beta}\left( \frac{m_*}{M_{\rm BH}} \right)^{1/3},
\label{eq:ecrit}
\end{equation}
below which all the stellar debris should remain gravitationally 
bound to the black hole. The critical eccentricity is evaluated to be 
$e_{\rm{crit}}=0.98$ for Model~4, whereas $e_{\rm{crit}}=0.996$ for Model~5.
For the eccentric TDEs, the orbital period of the most tightly bound orbit, $t_{\rm{min}}$, 
and the orbital period of the most loosely bound orbit, $t_{\rm{max}}$, 
are obtained 
by using Kepler's third law with $\epsilon=\Delta\epsilon\pm \epsilon_{\rm{orb}}$ and equation~(\ref{eq:eporb}). 
The duration time of mass fallback for eccentric TDEs with $e_*<e_{\rm{crit}}$ 
is thus predicted to be finite and can be written by 
$
\Delta{t}
=t_{\rm{max}} - t_{\rm{min}}
=(\pi/\sqrt{2})
(\Omega_{*}^{-1}/[\beta(1-e_{*})]^{3/2})
\left(
\left[ 
1/2 -
(1/\beta(1-e_{*})(m_*/M_{\rm{BH}})^{1/3} 
\right]^{-3/2}
-1
\right).
$
Evaluating this gives $\Delta{t}\approx207\Omega_{*}^{-1}$ for Model~5, 
whereas $\Delta{t}\rightarrow\infty$ for Model~4 in spite of smaller $t_{\rm{min}}$ 
than that of Models~1-3.

Figure~\ref{fig:mass-mdot} show 
differential mass distributions and their corresponding 
mass fallback rates in Models~1-5. While the differential 
mass distribution is shown in panel (a), the mass fallback rate
is shown in panel (b). 
In panel~(b), the horizontal solid line denotes the Eddington rate:
$\dot{M}_{\rm{Edd}}=(1/\eta)(L_{\rm{Edd}}/c^2)\simeq
2.2\times10^{-2}(\eta/0.1)^{-1}(M_{\rm{BH}}/10^6M_\odot)\,\,M_\odot\rm{yr^{-1}}$,
where $L_{\rm{Edd}}=4\pi GM_{\rm{BH}}m_{\rm{p}}c/\sigma_{\rm{T}}$ 
is the Eddington luminosity with $m_{\rm{p}}$ 
and $\sigma_{\rm{T}}$ denoting the proton mass and 
Thomson scattering cross section, respectively, 
and $\eta$ is the mass-to-energy conversion efficiency,
which is set to $0.1$ in the following discussion.

In panel~(a), the central peak of Model~1 is attributed 
to mass congregation, from the self-gravity of the stellar debris.
The energy spread corresponds to $\Delta\epsilon$ before and 
after the tidal disruption. The corresponding mass fallback rates 
are proportional to $t^{-5/3}$.
The slight deviation from time to the $-5/3$ power originates from the 
convexity around $\Delta\epsilon$ and the central peak rising from 
$0.2\Delta\epsilon$ to $0$ (see also \cite{lg09}).
Simulations of Models 2 and 3 have performed with the pseudo-Newtonian potential given by equation~(\ref{eq:wegg}).
Model~3 has the same simulation parameters as Model~2 except for $\beta=5$.
Since the potential is deeper as $\beta$ is higher, the re-congregation of the mass 
due to the self-gravity of the stellar debris is prevented. This leads to the mildly-sloped 
mass distribution, and therefore the peak of the mass fallback rate also smooths.

The mass is not distributed around zero but  
around $-\Delta\epsilon$ in Model~4, and around $-5\Delta\epsilon$ in Model~5. 
This is because the specific energy of initial stellar orbit is originally negative (see
equation~\ref{eq:eporb}).  
Clearly, most of mass in Model~4 is bounded by the negative shift of the center.  
The resultant energy spread is slightly larger than we analytically
expected. This suggests that the critical eccentricity is smaller than the value in equation~(\ref{eq:ecrit}).
In Model~5, all of mass is bounded and falls back to the black hole 
in a much shorter time than that of Models 1-3. As shown in panel~(b), the mass fallback rate of Model~5 is four orders of magnitude greater than the Eddington rate.

%
\section{Concluding remarks}
\label{sec:4}
%

We have performed 3D SPH simulations of tidal disruption processes for stars 
on bound orbits. Our main conclusions are summarized as follows:
\begin{enumerate}
\renewcommand{\theenumi}{\arabic{enumi}}
\item There is a critical orbital eccentricity below which all stellar debris falls back to the black hole.
The simulated critical eccentricity is slightly lower than expected from our analytical prediction. 
\item In an eccentric TDE with orbital eccentricity below the critical eccentricity, all the stellar debris falls back to the black hole in a much shorter time than that of the standard TDE. The resultant mass fallback rate substantially exceeds the Eddington rate and differs from the -5/3 power of time.
\end{enumerate}
The full details of this work can be seen in Ref. \cite{kh12}.


\end{document}